\documentclass{aa}

\usepackage{graphicx}
\usepackage{amsmath}
\usepackage{natbib}
\bibpunct{(}{)}{;}{a}{}{,}
\begin{document}

\title{Astrophysical bow shocks: An analytical solution for the hypersonic blunt body problem in the intergalactic medium}

\author{Michael M. Schulreich \and Dieter Breitschwerdt}
\institute{Zentrum f\"ur Astronomie und Astrophysik, Technische Universit\"at Berlin, Hardenbergstr.~36, D-10623 Berlin, Germany}

\mail{schulreich@astro.physik.tu-berlin.de}

\date{Received / Accepted}

\titlerunning{An analytical solution for the hypersonic blunt body problem in the IGM}
\authorrunning{M.M. Schulreich \& D. Breitschwerdt}

\abstract {} 
{Bow shock waves are a common feature of groups and clusters of galaxies since they are generated as a result of supersonic motion of
galaxies through the intergalactic medium. 
The goal of this work is to present an analytical solution technique for such astrophysical hypersonic blunt body problems. } 
{A method, developed by Schneider (1968, JFM, 31, 397) in the context of aeronautics, allows calculation of the galaxy's shape as long as the 
shape of the bow shock wave is known (so-called inverse method). In 
contrast to other analytical models, the solution is valid in the whole flow region (from the stagnation point up to the bow shock wings) 
and in particular takes into account velocity gradients along the streamlines. We compare our analytical results with two-dimensional 
hydrodynamical simulations carried out with an extended version of the {\small VH-1} hydrocode which is based on the piecewise parabolic 
method with a Lagrangian remap.}
{It is shown that the applied method accurately predicts the galaxy's shape and the fluid variables in the post-shock flow, 
thus saving a tremendous amount of computing time for future interpretations of similar objects. We also find that the method can be 
applied to arbitrary angles between the direction of the incoming flow and the axis of symmetry of the body. We emphasize that it is 
general enough to be applied to other astrophysical bow shocks, such as those on stellar and galactic scales.} 
{}
\maketitle

\section{Introduction}
\label{intro}

Astrophysical bow shocks are ubiquitous in the Universe and can be observed on all scales, from the Earth's bow shock through the heliospheric
(driven by the moving solar wind) and on to corresponding stellar wind bow shocks. On even larger scales, we see bow shocks of galaxies 
in the intergalactic medium (IGM) in groups and clusters, and possibly even in mergers of galaxy clusters \citep{Mar:02}. The main 
ingredients of the feature are a compressible medium and a body moving through it supersonically. This was already investigated
decades ago in aeronautical engineering in the context of supersonic aircrafts, and is commonly known as the supersonic blunt body 
problem. However in astronomy, with the exception of a few papers, analytical solutions have been scarce. For example, the paper of 
\cite{Can:98} considers the supersonic motion of a spherical body and treats the post-shock flow in a thin-shell approximation, which 
unfortunately restricts the applicability of the method to more realistic problems. The purpose of this paper is to introduce a more 
general analytical method, which allows calculating the stand-off distance of the shock and the complete post-shock flow. To 
demonstrate this, the analytic solutions will be compared to numerical two-dimensional simulations. It will be shown that the method even 
works reasonably well in the case of low Mach number shocks. As an important example of great astrophysical relevance, we analyse the bow 
shock of a galaxy moving supersonically through the IGM and show that the associated soft X-ray emission can be modelled fairly well.  

About half of the galaxies in the Universe are found in groups and clusters, which are large complexes of galaxies held together by the mutual 
gravitational attraction of their members, intergalactic gas, and, above all, dark matter. With the advent of imaging telescopes in X-ray 
astronomy, it became evident that galaxy clusters are intensive sources of X-ray radiation owing to the hot plasma located between the 
galaxies and trapped in the group's/cluster's potential well. Because of their low relative speeds, galaxies in groups and poor 
clusters affect each other and the surrounding IGM gravitationally stronger than the faster-moving cluster galaxies, leading to a 
variety of fascinating interaction or even merging processes. Galaxies dashing at super- or hypersonic velocities through the IGM produce 
wakes of gravitationally focused gas, lose mass due to ram pressure, thereby injecting metals into their gaseous environment. Finally, 
leading bow shock waves are generated ahead of the galaxies that change the state of the gas irreversibly and thus play 
an important role in the galaxies' structure and subsequent evolution \citep{Ste:99}.

\cite{Tri:03, Tri:05} have studied the complex X-ray emission of the compact galaxy group Stephan's Quintet (SQ) quite recently using 
\emph{Chandra} and \emph{XMM-Newton} observations. The prominent shock situated in the galaxy system has been resolved with \emph{Chandra} 
into a narrow north-south, somewhat clumpy structure between NGC 7318ab and NGC 7319, which is more sharply bounded on the west side (probably
due to a contact surface stabilized by a magnetic field that is indicated by radio continuum emission) and embedded in a more 
extended diffuse emission, which presumably represents preexisting IGM heated up by previous collisions (presumably NGC 7320c was 
involved). In the simplest scenario, as \cite{Tri:03, Tri:05} report, the shock results from the high-velocity collision of the gas-rich 
spiral galaxy NGC 7318b with previously stripped \ion{H}{i}  gas in SQ. For an upstream \ion{H}{i} temperature of 100\,K (and number 
density of about $6.5\times 10^{-3}$\,cm$^{-3}$), the velocity of the intruding galaxy is highly hypersonic ($\sim 1400\,$km\,s$^{-1}$), 
resulting in an upstream Mach number in the galaxy's rest frame of $M_{\infty}\simeq 930$. The gas inside the bow shock is then heated 
up to an (observed) energy of 0.5\,keV. To explain these low post-shock temperature, the authors suggest an oblique shock 
scenario with a shock inclination angle $\beta$ of about 30$^\circ$. Oblique shocks are a general feature of bow shocks, which are 
ultimately responsible for the observed IGM structures. The analytic calculation of general bow shocks, however, is fairly complicated. 
Previous attempts bear severe restrictions, like the spherical shape of the body and the thin-shell approximation of the model by 
\cite{Can:98}. On the other hand, bow shocks are ubiquitous phenomena in the intergalactic and interstellar medium (ISM), and an analytic 
description like the one given in the present paper is therefore most desirable.

The paper is organized as follows. In Section~\ref{method} the analytic solution for the blunt body problem is presented. In 
Section~\ref{results}, this method is applied to Stephan's Quintet, and the results are also compared to numerical two-dimensional 
simulations. Section~\ref{conc} closes the paper with our conclusions.

\section{Solution for the hypersonic blunt body problem}
\label{method}
\begin{figure}
\resizebox{\hsize}{!}{\includegraphics{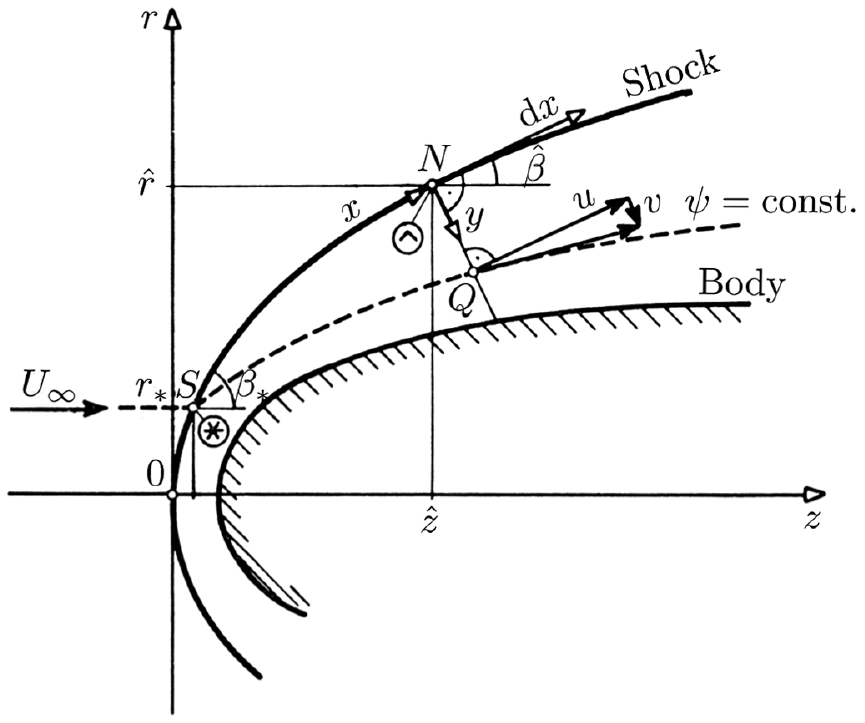}}
\caption{Shock-oriented coordinate system of boundary-layer type, where $z$ and $r$ are Cartesian coordinates for plane flow, $x$ and $y$ 
are the distances along the shock surface and normal to it, with $u$ and $v$ as the corresponding velocity components, $U_\infty$  is the 
free-stream velocity, and $\psi$ is the stream function. Flow quantities immediately behind the shock in the point $N$ are denoted by a 
hat (~$\hat{}$~), and in the point $S$ by an asterisk ($*$) \citep[adapted from][]{Sch:68}.}
\label{im:coordsys}
\end{figure}
A body moving supersonically through a gaseous medium will generate a shock wave, which is generally detached and curved. It appears to 
be planar at the body's nose and becomes progressively more oblique sideways, until it decays into a sonic wave at large distances. As a 
result, this single shock wave represents all possible oblique shock solutions for the given Mach number $M_\infty$ with the wave angle 
ranging from $\beta=\pi/2$ to $\beta=\alpha$, where $\alpha=\arcsin(1/M)$ is the so-called Mach angle, i.e.~the angle forming the cone to 
which small disturbances in a supersonic flow are confined. The blunt body shock layer, the volume between the body and the shock, is a mixed 
subsonic-supersonic flow, where the subsonic and supersonic regions are divided by sonic lines, i.e~the loci of points at which the 
downstream Mach number is unity. Behind the normal, and nearly normal portions of the shock wave, the flow is subsonic, whereas behind the 
more oblique portions of the shock wave the flow is supersonic. It is precisely this hybrid nature of the flow that makes the blunt body 
problem extremely challenging mathematically, since the governing highly nonlinear partial differential equations of hydrodynamics 
are of elliptic type in the subsonic region and of hyperbolic type in the supersonic region \citep{And:06}. 

The analysis developed by \cite{Sch:68} allows a very elegant treatment of the inviscid hypersonic blunt body problem. It is an 
\emph{inverse} method, which means that the shock wave shape is assumed, and both the body shape,  which supports the assumed shock, and 
the flow field between shock and body (i.e.~the shock layer) are calculated. The fundamental advantage of this method over the 
ones by other authors is its uniform validity in the whole flow field (from the stagnation region up to large distances from the 
projectile nose). Until now this method has been adopted in aerospace engineering, in particular for solving the reentry problem of space 
probes, space shuttles, etc., in planetary or terrestrial atmosphere. However, to the best of our knowledge, this paper demonstrates its first 
application to an astrophysical problem.

We consider the plane or axisymmetric flow around a body and introduce a shock-oriented curvilinear coordinate system of 
boundary-layer type, where $x$ is the distance along the shock surface in the plane formed by the shock normal and the direction of the 
uniform fluid flow, and $y$ is the distance normal to the shock surface (Fig.~\ref{im:coordsys}). The corresponding velocity components 
are denoted by $u$ and $v$, and $z$ and $r$ are the Cartesian coordinates for plane flow or the cylindrical coordinates for axisymmetric 
flow. The $z$-axis may be parallel to the direction of the incident flow. We emphasize that this is only assumed for convenience, but is 
no general restriction (see Section \ref{conc}). It can be deduced from Fig.~\ref{im:coordsys} that
\begin{align}
z & = \hat{z}+y\,\sin{\hat{\beta}}\,,\label{eq:trafoz}\\
r & = \hat{r}-y\,\cos{\hat{\beta}}\,, \label{eq:trafor}
\end{align}
where $\hat{\beta}$ is the shock inclination angle in the point $N(\hat{z},\hat{r})$, with the shock-normal through $Q$ intersecting the 
shock surface. Moreover, $S$ is the point where the streamline through $Q$ crosses the shock wave. The flow quantities immediately behind 
the shock in the point $N$ are denoted by a hat (~$\hat{}$~), and in the point $S$ by an asterisk ($*$). Undisturbed flow quantities 
far upstream are denoted by the subscript $\infty$. 

Since the functions $\hat{z}(x)$ and $\hat{r}(x)$ are known for a given shock shape, equations (\ref{eq:trafoz}) and (\ref{eq:trafor}) may 
be used to calculate the coordinates $z$ and $r$ of a point $Q$ from its coordinates $x$ and $y$. The curvature of the shock contour in 
the point $N$ is denoted by $\hat{\kappa}(x)$, defined as positive when the surface is concave on the side of positive $y$ 
(cf.~Fig.~\ref{im:coordsys}). The curvature of any of the other surfaces of constant $y$ is $\mathcal{H}^{-1}\hat{\kappa}$, where    
\begin{equation}
\mathcal{H}=1-\hat{\kappa} y>0\,.
\end{equation}
The metric for this coordinate system is \citep[cf.][]{Hay:66}
\begin{equation}
\text{d}s^2=\mathcal{H}^2\text{d}x^2+\text{d}y^2\,.
\end{equation}
Thus, the governing hydrodynamical steady-state equations for mass, momentum, and energy become
\begin{align}
\frac{\partial r^j\rho u}{\partial x}+\frac{\partial \mathcal{H}r^j\rho v}{\partial y}&=0\,,\\
u\frac{\partial u}{\partial x}+\mathcal{H}v\frac{\partial u}{\partial y}-\hat{\kappa} uv+\frac{1}{\rho}\frac{\partial P}{\partial x}&=0\,,
\\
u\frac{\partial v}{\partial x}+\mathcal{H}v\frac{\partial v}{\partial y}+\hat{\kappa} u^2+\frac{\mathcal{H}}{\rho}\frac{\partial P}
{\partial y}&=0\,,\\
u\frac{\partial S}{\partial x}+\mathcal{H}v\frac{\partial S}{\partial y}&=0\,,
\end{align}
where $\rho$ is the fluid density, $P$ the fluid pressure, $S$ the entropy, $r(x,y)$ the distance from the axis. The parameter $j$ is 0 or 
1 for plane or axisymmetric flow, respectively. We used the entropy equation in place of the energy equation. We now want
to make a von Mises transformation such that the independent variables $(x,y)$ are replaced by $(\bar{x}=x,\psi)$, where $\psi$ is a 
stream function defined in terms of its partial derivatives
\begin{align}
\frac{\partial \psi}{\partial x}&=(1-\hat{\kappa}y)r^j\rho v\,,\\
\frac{\partial \psi}{\partial y}&=-r^j\rho u\,.
\end{align}
The stream function is constant along a streamline and represents the mass flow between the streamline $\psi=\text{const.}$ and the surface 
of the projectile, per unit depth for plane flows, and per unit azimuthal angle (in radians) for axisymmetric flows; i.e., $\psi$ satisfies 
the continuity equation. In the point $N$, the stream function thus is simply
\begin{equation}
\hat{\psi}=\rho_\infty U_\infty \frac{\hat{r}^{1+j}}{1+j}\,.
\label{eq:psihat}
\end{equation}
In contrast, $\psi$ is connected with the coordinate $r_*$ of the point $S$ by 
\begin{equation}
\psi=\rho_\infty U_\infty \frac{r_*^{1+j}}{1+j}\,.
\end{equation}
In the new coordinate system with the variable $y$ eliminated, the equations of energy, entropy, and momentum conservation read as
\begin{equation}
u^2+v^2+2h=u_*^2+v_*^2+2h_*=\text{const.}\,,
\label{eq:schnenergy}
\end{equation}

\begin{equation}
\frac{\partial S}{\partial \bar{x}}=0,\;\;\;\;\;\;\text{or}\;\;\;\;\;\;S=S_*(\psi)\,,
\label{eq:entbound}
\end{equation}
\begin{equation}
u\frac{\partial u}{\partial \bar{x}}+v\frac{\partial v}{\partial \bar{x}}+\frac{1}{\rho}\frac{\partial P}{\partial \bar{x}}=0\,,
\label{eq:schnmom1}
\end{equation}
\begin{equation}
(1-\hat{\kappa}y)\left(1-j\frac{y}{\hat{r}}\cos\hat{\beta}\right)\hat{r}^j\frac{\partial P}{\partial \psi}=\hat{\kappa}u+\frac{\partial v}
{\partial \bar{x}}\label{eq:schnmom2}\,,
\end{equation}
where $h$ is the specific enthalpy.

The variable $y$ is a dependent variable after the von Mises transformation, and it obeys the equations
\begin{align}
\frac{\partial y}{\partial \bar{x}}&=(1-\hat{\kappa}y)\frac{v}{u}\label{eq:diffy1}\,,\\
\frac{\partial y}{\partial \psi}&=-\frac{1}{\{1-j(y/\hat{r})\cos\hat{\beta}\}\hat{r}^j\rho u} \label{eq:diffy}\,.
\end{align}

At this point it is useful to introduce the Landau symbol $\mathcal{O}$, which describes an asymptotic upper bound for the magnitude of a 
function in terms of another, usually simpler function; e.g., $f(x)=\mathcal{O}(g(x))$ means that $|f(x)|$ is not very large in comparison 
with $|g(x)|$. The flow quantities immediately behind the shock may be obtained from the Rankine-Hugoniot jump conditions in terms of the 
inverse compression ratio across the shock $\hat{\chi}=\rho_\infty/\hat{\rho}$. They are 
\begin{align}
\hat{u}&=U_\infty \cos\hat{\beta}\,\label{eq:uhat},\\
\hat{v}&=U_\infty \hat{\chi}\sin\hat{\beta}\,\label{eq:vhat},\\
\hat{P}&=P_\infty+\rho_\infty U_\infty^2(1-\hat{\chi})\sin^2\hat{\beta}\,\label{eq:p_hat},\\
\hat{h}&=h_\infty+\frac{1}{2}U_\infty^2(1-\hat{\chi}^2)\sin^2\hat{\beta}\label{eq:enthbound}\,.
\end{align}
These equations maintain their validity if the hats are replaced by asterisks. The pressure within the shock layer is given by the 
\emph{Newton-Busemann pressure law} \citep[see e.g.][]{Hay:66}, a simplified form of the momentum equation (\ref{eq:schnmom2}), which 
reads as
\begin{equation}
P=\hat{P}-\frac{\hat{\kappa}}{\hat{r}^j}\int_\psi^{\hat{\psi}}u\,\text{d}\psi'\,.
\label{eq:newbuse0}
\end{equation}

The following method is based on two main assumptions. First, it is assumed that the inverse compression ratio across the shock is very 
small; i.e.,
\begin{equation}
\hat{\chi}=\frac{\rho_\infty}{\hat{\rho}}\ll1 \;\;\;\;\;\;\text{and}\;\;\;\;\;\; \chi_*=\frac{\rho_\infty}{\rho_*}=\mathcal{O}(\hat{\chi})
\,.
\label{eq:ass1}
\end{equation}
Second, the pressure at the point $Q$ of the disturbed flow field should not be much smaller than the pressure immediately behind 
the shock in the intersection point of the shock surface with its normal through the point $Q$; i.e.,
\begin{equation}
\frac{\hat{P}}{P}=\mathcal{O}(1)\;\;\;\;\;\; (\text{on } x=\text{const, } y>0)\,.
\label{eq:ass2}
\end{equation}
We follow \cite{Sch:68} to express the pressure in an arbitrary point $Q$ approximately by (see Appendix)
\begin{equation}
P=\hat{P}-\frac{\hat{\kappa}}{\hat{r}^j}\int_\psi^{\hat{\psi}}\{u_*^2+2[h_*-h(\hat{P},S_*)]\}^{1/2}\text{d}\psi'\,. 
\label{eq:newbuseneu}
\end{equation}
It may be noted that \emph{all} quantities on the equation's right-hand side are given by the boundary conditions at the shock or by the 
equation of state. Terms coming from the excluded portions of the flow field -- namely the stagnation region, as well as the region near 
the stagnation region where $u\ll \hat{u}$ -- are of the order of $\hat{\chi}$ and therefore contribute only negligibly to the 
integral in equation (\ref{eq:newbuseneu}). Consequently, the \emph{whole} gas-dynamic state is known in the streamline coordinate system 
($\bar{x},\psi$) by evaluating $S=S_*(\psi)$ and $P$ from equation (\ref{eq:newbuseneu}). 

The actual location of the body in space can be determined by solving the differential equation (\ref{eq:diffy}) via separation of 
variables giving the distance from the shock surface $y$ as a function of $\bar{x}$ and $\psi$;
\begin{equation}
y\left(1-\frac{j\cos\hat{\beta}} {2\hat{r}} y\right)=\frac{1} {\hat{r}^j} \int_\psi^{\hat{\psi}}\frac{\text{d}\psi'}{\rho u}\,.
\label{eq:schny}
\end{equation}
Neglecting errors of $\mathcal{O}(\hat{\chi})$ we may replace the velocity component in $x$-direction by
\begin{equation}
u^2=u_*^2+2[h_*-h(P,S_*)]+\dots\,,
\end{equation}
which follows from the energy equation (\ref{eq:schnenergy}). The integral in equation (\ref{eq:schny}) is then
\begin{equation}
Y=\int_\psi^{\hat{\psi}}\frac{\text{d}\psi'}{\rho(P,S_*)\{u_*^2+2[h_*-h(P,S_*)]\}^{1/2}}\,.
\label{eq:ypsilon}
\end{equation}
Solving the quadratic equation in $y$ on the left-hand side of equation (\ref{eq:schny}), we have to distinguish between plane ($j=0$) and 
axisymmetric ($j=1$) flows. Thus we have
\begin{align}
&\text{for} \hspace{0.2cm} j=0:\hspace{0.2cm} y=Y\,;\\
&\text{for} \hspace{0.2cm} j=1:\hspace{0.2cm} y=\frac{\hat{r}}{\cos\hat{\beta}}\left[1-\left(1-\frac{2Y\cos\hat{\beta}}{\hat{r}^2}\right)^
{1/2}\right]\,.
\end{align}
To finally give these results in the convenient coordinates $x$ and $y$, the transformations (\ref{eq:trafoz}) and 
(\ref{eq:trafor}) have to be carried out.

In the astrophysical context, the case of a perfect gas with constant specific heats is of great interest. The inverse compression ratio is 
then given by \citep[see e.g.][]{Lan:04}
\begin{equation}
\chi_*=\frac{\gamma-1}{\gamma+1}+\frac{2}{(\gamma+1)M_\infty^2\sin^2\beta_*}\,.
\label{eq:chistar}
\end{equation}
We point out that an analogous relation is valid for $\hat{\chi}$, if all asterisks in equation (\ref{eq:chistar}) are replaced by 
hats.
 
Using the shock conditions (\ref{eq:uhat})--(\ref{eq:enthbound}), together with equation (\ref{eq:chistar}), the two integrals 
(\ref{eq:newbuseneu}) and (\ref{eq:ypsilon}), which have to be evaluated, become
\begin{equation}
\begin{split}
P =\hat{P}-\frac{U_\infty\hat{\kappa}}{\hat{r}^j}\int_\psi^{\hat{\psi}}\limits &\Bigg\{ \cos^2\beta_*+  \left[\frac{2}{(\gamma-1)M_
\infty^2}+\sin^2\beta_*  \right] \\
&\times  \left[  1-  \left(\frac{\sin^2\hat{\beta}}{\sin^2\beta_*}  \right)^{\frac{\gamma-1}{\gamma}}  \right]\Bigg\}^{1/2}\text{d}
\psi'\,,
\label{eq:perfectpressure}
\end{split}
\end{equation}
\begin{equation}
\begin{split}
Y&=\frac{1}{\rho_\infty U_\infty}\int_\psi^{\hat{\psi}}\limits \chi_* (\hat{P}\sin^2\beta_*/P\sin^2\hat{\beta})^{1/\gamma} \\
&\times\Bigg\{\cos^2\beta_*+\left[\frac{2}{(\gamma-1)M_\infty^2}+\sin^2\beta_*\right]\\
&\times\left[1-\left(\frac{P\sin^2\hat{\beta}}{\hat{P}\sin^2\beta_*}\right)^{\frac{\gamma-1}{\gamma}}\right]\Bigg\}^{-1/2} \text{d}
\psi'\,.
\end{split}
\label{eq:perfectY}
\end{equation}
The curvature of a curve in space can be calculated by the well-known formula
\begin{equation}
\kappa=\frac{\left|\frac{\text{d}^2r}{\text{d}z^2}\right|}{\left[1+\left(\frac{\text{d}r}{\text{d}z}\right)^2\right]^{3/2}}\,.
\end{equation}
In addition, the following simple relations for the the shock inclination angle turn out to be quite useful:
\begin{align}
\tan\beta &=\frac{\text{d}r}{\text{d}z}=:q\,,\\
\sin^2\beta &= \frac{q^2}{1+q^2}\,,\\
\cos^2\beta &= \frac{1}{1+q^2}\,.
\end{align}
The density for a perfect gas is represented by the quantity
\begin{equation}
\begin{split}
\rho=\rho_* \left(\frac{P}{P_*}\right)^{1/\gamma}&=\left(\frac{\gamma-1}{\gamma+1}+\frac{2}{(\gamma+1)M_\infty^2\sin^2\beta_*}\right)^
{-1}\\
&\times\rho_\infty\left(\frac{P}{\hat{P}}\frac{\sin^2\hat{\beta}}{\sin^2\beta_*}\right)^{1/\gamma}\,.
\end{split}
\end{equation}
For simplicity we introduce dimensionless units; e.g.~the normalized stream function then is $\Psi=\psi/\rho_\infty U_\infty L^{j+1}$, 
where $L$ is a characteristic length. 

On the surface of the body we need $\psi=0$, so the pressure on the body surface $P_\text{b}(x)$, as well as the shock layer thickness $
\Delta(x)$, can be obtained by replacing the lower limits in equations (\ref{eq:perfectpressure}) and (\ref{eq:perfectY}) by zero:

\begin{equation}
\begin{split}
\frac{P_\text{b}}{\rho_\infty U_\infty^2}&=\frac{1}{\gamma M_\infty^2}+(1-\hat{\chi})\sin^2\hat{\beta}\\
&-\frac{\hat{\kappa}}{\hat{r}^j}\int_0^{\hat{\Psi}}\limits \Bigg\{ \cos^2\beta_*+\left[\frac{2}{(\gamma-1)M_\infty^2}+\sin^2\beta_*  
\right] \\ 
&\times\left[  1-  \left(\frac{\sin^2\hat{\beta}}{\sin^2\beta_*}  \right)^{\frac{\gamma-1}{\gamma}}  \right]\Bigg\}^{1/2}\text{d}\Psi\,,      
\end{split} 
\label{eq:int1}                                                                                                              
\end{equation}
where the relation
\begin{equation}
\begin{split}
\frac{\hat{P}}{\rho_\infty U_\infty^2}&=\frac{P_\infty}{\rho_\infty U_\infty^2}+(1-\hat{\chi})\sin^2\hat{\beta}\\
                                                       &=\frac{1}{\gamma M_\infty^2}+(1-\hat{\chi})\sin^2\hat{\beta}
\end{split}
\end{equation}
has been used. Furthermore, we get
\begin{align}
\text{for} \hspace{0.2cm}j=0&:\hspace{0.1cm} \Delta=Y\bigg|_{\psi=0}\,,\label{eq:delta1}\\
\text{for} \hspace{0.2cm}j=1&:\hspace{0.1cm} \Delta=\frac{\hat{r}}{\cos\hat{\beta}}\left[1-\left(1-\frac{2Y|_{\psi=0}\cos\hat{\beta}}{\hat
{r}^2}\right)^{1/2}\right],\label{eq:delta2}
\end{align}
with
\begin{equation}
\begin{split}
Y\bigg|_{\psi=0}&=\int_0^{\hat{\Psi}}\limits\chi_* (\hat{P}\sin^2\beta_*/P\sin^2\hat{\beta})^{1/\gamma} \\
&\times\Bigg\{\cos^2\beta_*+\left[\frac{2}{(\gamma-1)M_\infty^2}+\sin^2\beta_*\right] \\
&\times \left[1-\left(\frac{P\sin^2\hat{\beta}}{\hat{P}\sin^2\beta_*}\right)^{\frac{\gamma-1}{\gamma}}\right]\Bigg\}^{-1/2}\text{d}\Psi\,.
\end{split}
\label{eq:int2}
\end{equation}

As soon as the bow shock wave is parametrized, the gas-dynamical quantities can be computed by evaluating two integrals (namely equations 
(\ref{eq:int1}) and (\ref{eq:delta1}) or (\ref{eq:delta2}) and (\ref{eq:int2}), together  with the boundary values (\ref{eq:entbound}) and 
(\ref{eq:uhat})--(\ref{eq:enthbound})). 

\section{Results}
\label{results}

\subsection{Validation}

\begin{figure}
\resizebox{\hsize}{!}{\includegraphics{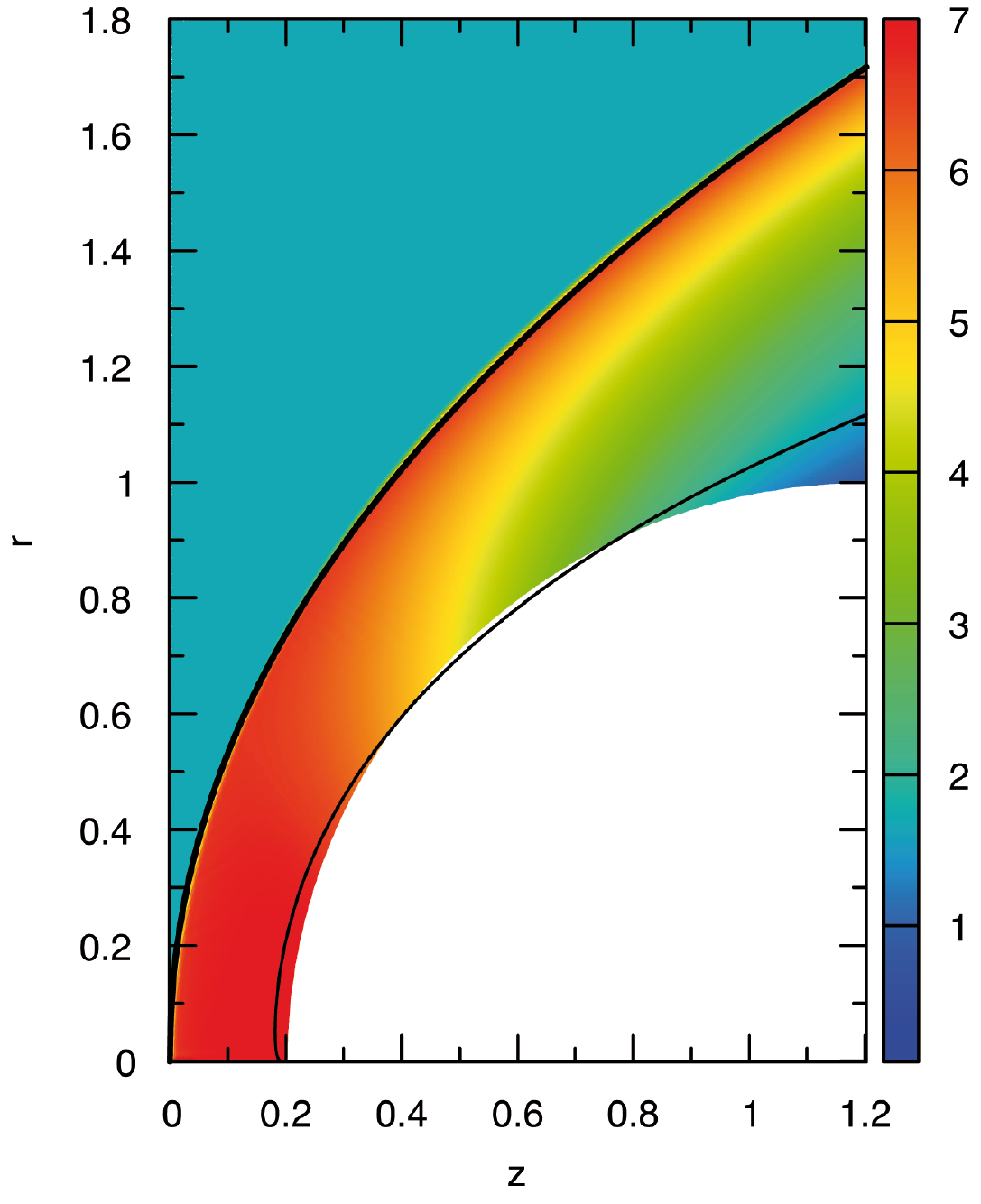}}
\caption{Comparison between the numerical and analytical solution of a bow shock wave generated by a hypersonic flow ($M_{\infty}=500$) 
hitting the solid unit sphere. A perfect monoatomic gas is assumed ($\gamma=5/3$). The analytical body prediction is represented by the 
thin solid line. The bow shock wave (thick solid line) is parametrized by $\hat{r}=(2.620\cdot \hat{z})^{0.471}$ with $j=1$. The 
(dimensionless) density profile shown is colour-coded, with red the high and blue the low density.}
\label{im:solid}
\end{figure}

The applicability of the analytical model for an astrophysical context is tested by simulating the bow shock wave induced by the 
interaction of an uniform hypersonic flow with the solid unit sphere. For the sake of simplicity we have chosen 
$M_{\infty}\equiv U_{\infty}=500$, which is ensured by setting $\rho_{\infty}\equiv\gamma=5/3$ and $P_{\infty}=1$. A simulation snapshot 
overlaid by the analytical solution is depicted in Fig.~\ref{im:solid}. At the stage shown here, the simulation was evolved long enough 
($t=0.2$) so that the bow shock wave has already become stationary for some time. The modulus of the percentage error of the stagnation 
distance prediction is $\sim 6.2\,\%$, and the modulus of the maximum percentage error of the obstacle estimation is $\sim 11.6\,\%$. These 
errors may arise from two sources; first, the analytical method is based on the assumption that the density ratio $\chi$ across the shock 
is very small. Since we consider here a perfect monoatomic gas with $\gamma=5/3$, an error that can reach the $25\,\%$ is introduced 
(cf.~Eq.~(\ref{eq:chistar})). However, for the astrophysically more relevant case of a cooling strong shock, the compression ratio of four 
only represents a lower limit. Thus, when the cooling time proceeds, relation (\ref{eq:ass1}) becomes progressively more precise. Second, the 
hypersonic gas flow around a sphere features a complex wake that contains inter alia a second shock wave as a result of overexpansion 
with following compression. This explains especially the increasing deviation of the analytical solution from the numerical result at (or 
in the vicinity of) the sphere's equator (Schneider, priv.~comm.).

\subsection{Application to observations}
\label{sec:obs}
\begin{figure}
\resizebox{\hsize}{!}{\includegraphics{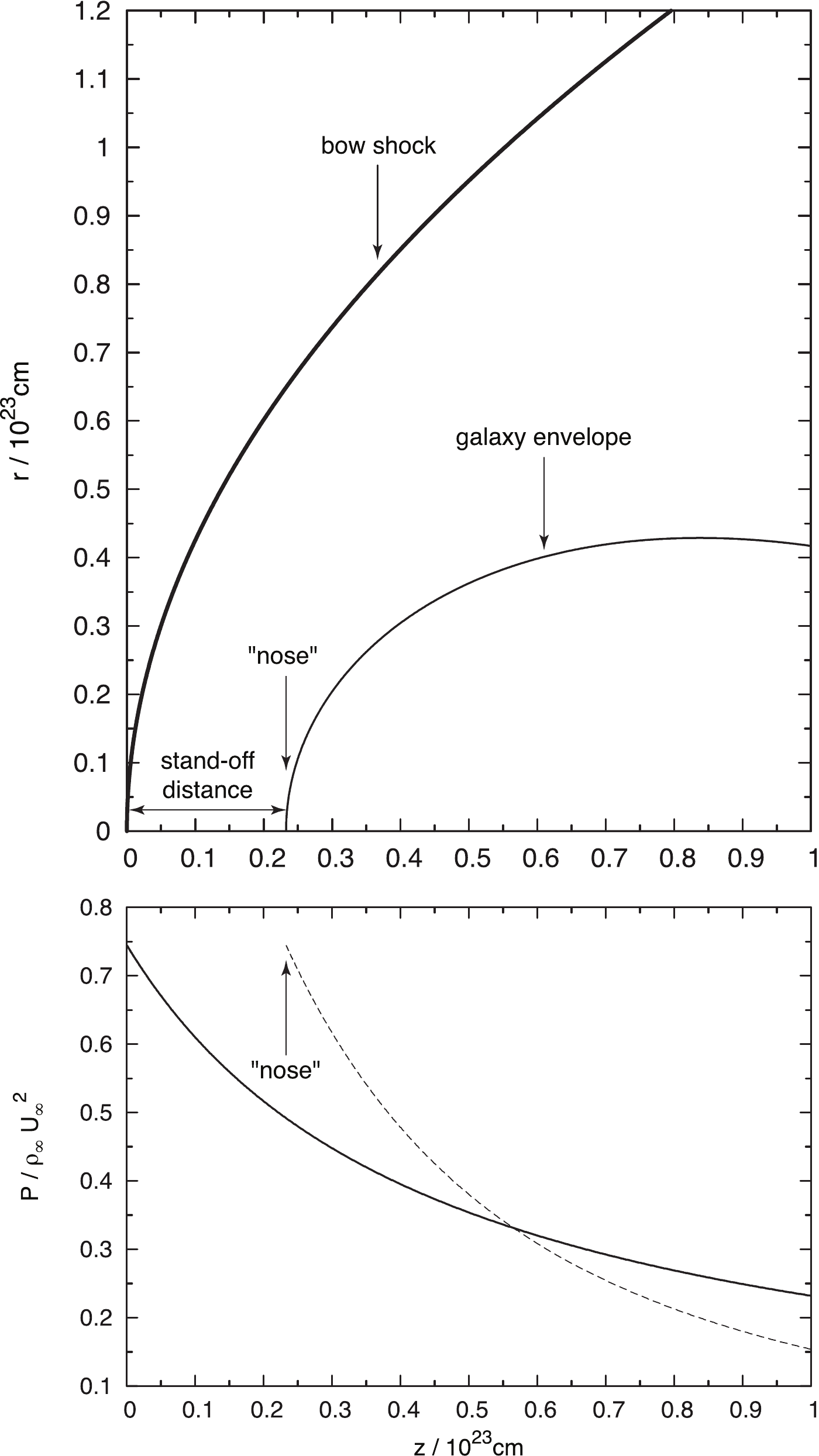}}
\caption{\emph{Upper panel.} Shape of the SQ galaxy NGC 7318b (thin line) and the bow shock (thick line) at $t=130\,$Myr ($j=0$, 
$\hat{r}^2=1.81\,\hat{z}$). A perfect gas with $\gamma=5/3$ and an upstream Mach number of $M_\infty=930$ have been assumed. 
\emph{Lower panel.} Normalized pressure $P/\rho_{\infty}U_{\infty}^2$ along the galaxy surface (i.e.~along the body streamline) (dashed line) 
and directly behind the bow shock wave (solid line).}
\label{im:analyt}
\end{figure}

\begin{figure*}
\resizebox{\hsize}{!}{\includegraphics{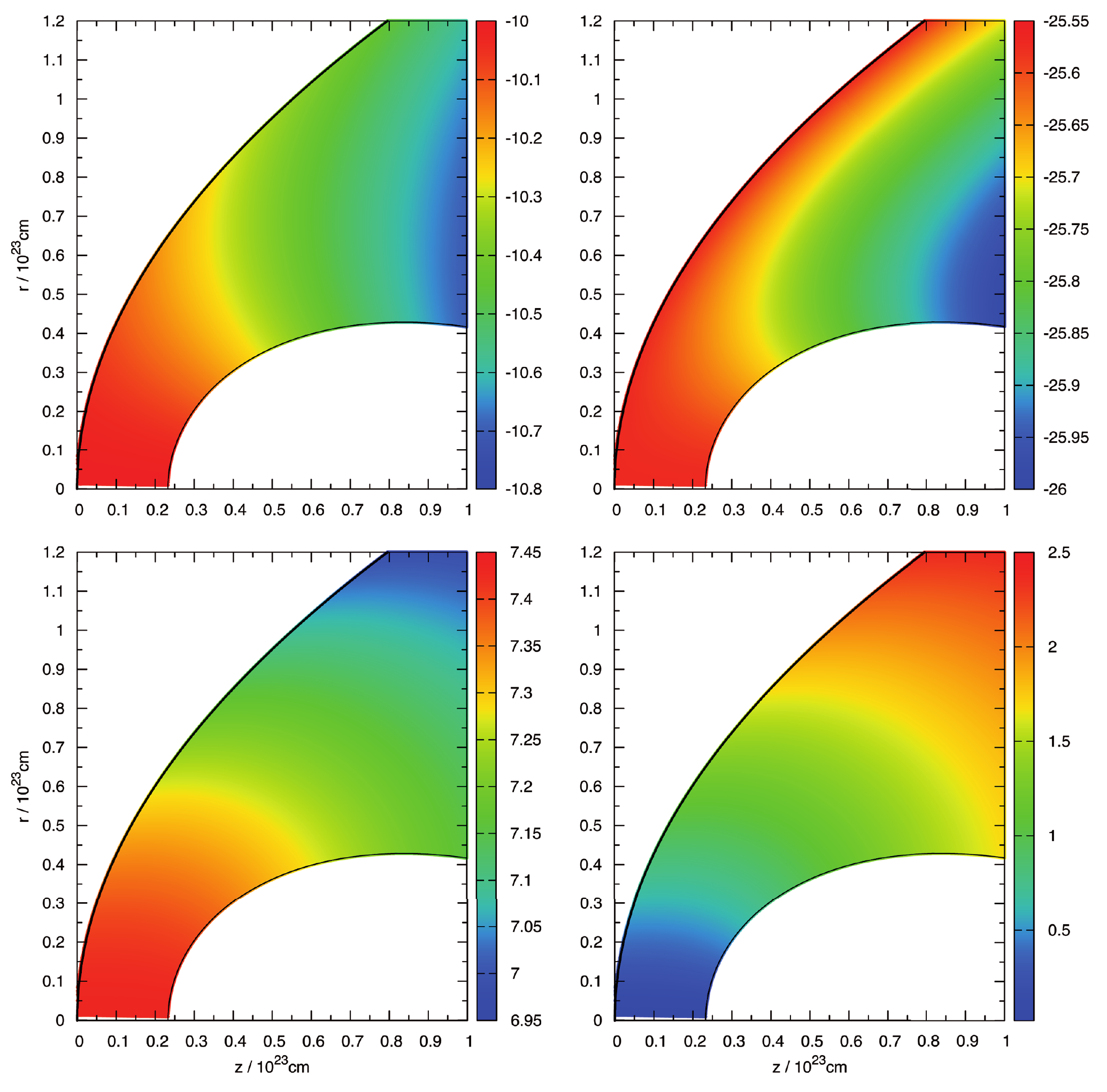}}
\caption{Analytically derived colour-coded maps of several flow quantities, namely log pressure $P$ in dyn\,cm$^{-2}$ (\emph{upper left 
panel}), log density $\rho$ in g\,cm$^{-3}$ (\emph{upper right panel}), log temperature $T$ in K (\emph{lower left panel}), and the 
velocity ratio $u/c_\text{s}\simeq M$ (\emph{lower right panel}) for the SQ galaxy NGC 7318b at the time $t=130\,$Myr ($j=0$, 
$\hat{r}^2=1.81\,\hat{z}$). A perfect gas with $\gamma=5/3$ and an upstream Mach number of $M_\infty=930$ have been assumed.}  
\label{im:flowfield}
\end{figure*}

To model the scenario in SQ, where a large-scale bow shock wave is believed to be formed by the gas-rich spiral galaxy NGC 7318b 
as it plunges through 100\,K H\,{\small I} gas \citep{Tri:03}, the upstream Mach number has chosen to be $M_\infty=930$. We assumed a 
plane \linebreak($j=0$) supersonic flow of perfect monoatomic gas ($\gamma=5/3$) and parametrized the bow shock time-dependently (e.g. by 
$\hat{r}^2=1.81\,\hat{z}$ for $t=130$\,Myr after the start of the interaction). Results are plotted in Figs.~\ref{im:analyt} and 
\ref{im:flowfield}.

The standoff distance of the shock from the nose of the projectile (in this case from the leading edge of the galaxy NGC 7318b of SQ) can 
be immediately determined to be $\Delta_0\simeq 2.3\times 10^{22}\,$cm ($\simeq 7.4$\,kpc). In Fig.~\ref{im:flowfield} analytically 
derived maps of several fluid variables for the galaxy-IGM interaction process in SQ are shown, namely pressure $P$ (upper left panel), 
density $\rho$ (upper right panel), temperature $T$ (lower left panel), and the approximate Mach number (since we have ignored $v$ in the 
derivation, which is the velocity component in $y$-direction) $u/c_\text{s}\simeq M$ (lower right panel), where $c_\text{s}$ is the local 
downstream velocity of sound. Since values of $\hat{z}$ that are too low lead to a singularity in the (numerical) integration, the plane and 
near plane post-shock region are missing. We have written a Fortran program based on the presented method that allows the hypersonic 
blunt body problem to be solved for the flow quantities in the whole post-shock flow field. Required input variables are the parametrized bow 
shock $\hat{r}(\hat{z})$, the flow geometry factor $j$, the upstream Mach number $M_\infty$, and the adiabatic index of the gas $\gamma$. 
Furthermore, upstream  density, length scale, and the velocity of the obstacle (or flow, respectively) have to be known to allow an easy 
comparison to observational data. This tool is available online\footnote{\texttt{http://astro.physik.tu-berlin.de/downloads}}.

To test our analytical results we performed several two-dimensional numerical hydrodynamical simulations. We used the 
{\small VH-1} hydrocode as a basis, which was written by the Numerical Astrophysics Group at the Virginia Institute for Theoretical 
Astrophysics \citep{Blo:90}. The code is based on a third-order accurate extension of the grid-based Godunov scheme, namely a Lagrangian-remap version of the piecewise parabolic method (PPMLR), and features thus good shock capturing. Optically thin radiative cooling via 
operator splitting has been included. Each numerical simulation has been carried out on an uniform $2400\times 1200$ Cartesian grid, with 
a computational domain of $4.8\times 10^{23}\,$cm by $2.4 \times 10^{23}\,$cm ($\sim 156\times 78\,$kpc). This results in a grid 
resolution of $\Delta x=\Delta y = 2\times 10^{20}\,$cm ($\sim 65\,$pc). Runs with other grid sizes showed that the output does not 
depend on resolution. The only difference is that small-scale instabilities and fragmentations of stripped material cannot be resolved on 
coarser grids. ``Outflow'' boundary conditions have been used along the top, the bottom, and on the right of the grid, while the left 
boundary is ``inflow''. The galaxy itself was modelled by a simple stratified gaseous ellipse. Each computation begins with the galaxy 
placed face-on at rest in an uniform supersonic stream of intergalactic gas, corresponding to the situation of the galaxy rushing through 
the IGM at constant speed.

\begin{figure}
\resizebox{\hsize}{!}{\includegraphics{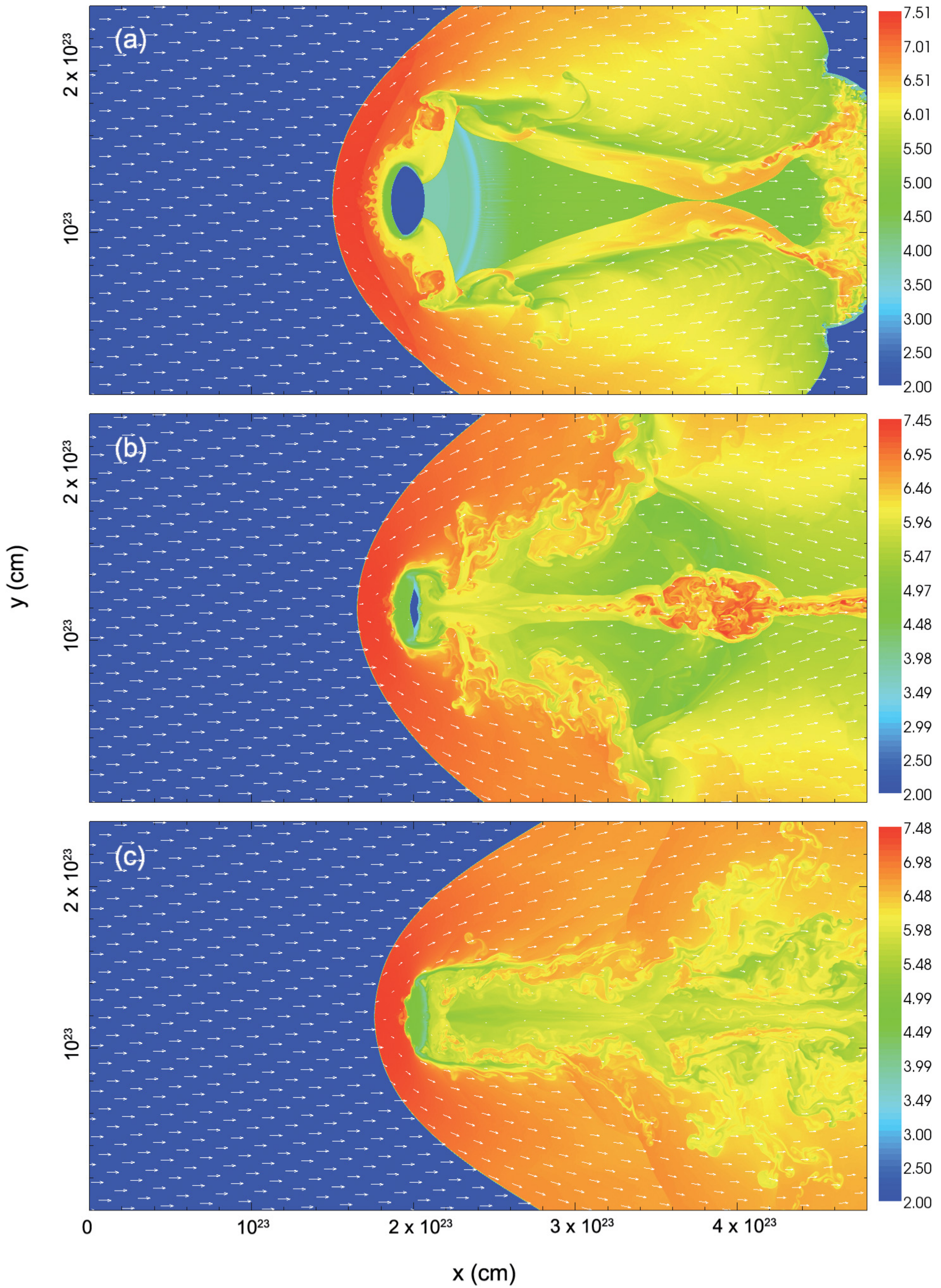}}
\caption{Colour-coded log temperature map (in K) overlaid with the velocity field of the SQ simulation at the times (a) $t= 60$\,Myr, (b) 
$t= 130$\,Myr, and (c) $t= 200\,$Myr. The vectors drawn show the direction and magnitude of the velocity field.}
\label{im:sim}
\end{figure}

Figure \ref{im:sim} gives an impression of our simulations. In this example, we tried to reconstruct the conditions that presumably occur in 
SQ. The ISM of the intruding galaxy NGC 7318b (whose semi-axes have been taken from the 
NED\footnote{\texttt{http://nedwww.ipac.caltech.edu/}}) consists in our simple model of three equidistant layers, which are in 
pressure equilibrium; namely, an \ion{H}{i} core ($\rho_{\text{core}}=6.2\times 10^{-24}\,$g\,cm$^{-3}$, $T_{\text{core}}=100\,$K), a 
Lockman layer ($\rho_{\text{Lock}}=1.0\times 10^{-25}\,$g\,cm$^{-3}$, $T_{\text{Lock}}=6\,000\,$K), and a Reynolds layer 
($\rho_{\text{Rey}}=6.9\times 10^{-26}\,$g\,cm$^{-3}$, $T_{\text{Rey}}=8\,000\,$K). The IGM ($\rho_{\text{IGM}}=6.7\times 10^{-27}\,$g\,cm
$^{-3}$, $T_{\text{IGM}}=100\,$K) flows from left to right at a constant hypersonic speed ($M_\infty=930$). Solar metallicity is assumed 
in all the considered media. In Figure \ref{im:sim} the colour-coded log temperature map (in K) with the superimposed velocity field is 
visualized at the times 60, 130, and 200\,Myr after the start of the simulation. Owing to the reasonably 
good numerical resolution of the simulation the presented snapshots allow tracking of a multitude of hydrodynamical effects (such as a prominent bow shock wave, 
Kelvin-Helmholtz instabilities, Rayleigh-Taylor instabilities, gas compression in the course of radiative cooling, a turbulent Karman wake, etc.) 
that lead to a subsequent evolution of the galaxy's ISM. Please note that more sophisticated \emph{magneto}hydrodynamic simulations with 
an underlying galaxy dark matter halo (which are beyond the scope of this paper) would be required to study a possible reduction or even 
suppression of fluid dynamical instabilities. However, at high Mach numbers, as they occur in the considered galaxy-IGM interactions, the 
velocity potential (kinetic energy density) $\mathcal{V} \sim u^{2}/2$ (with $u \sim1400$\,km/s) is at least five times higher 
than the gravitational potential (including dark matter) $\Phi\sim GM/R_{\text{min}}$ (with $M \sim 2\times 10^{45}$\,g, and a minimum 
distance from the galaxy's centre of $R_{\text{min}} \sim 7\times 10^{22}$\,cm). Thus, the bow shock, as well as the shock layer, are 
unlikely to be influenced significantly by gravity in general and dark matter in particular.

The calculated post-shock temperature at the intersection between the galaxy's axis of symmetry and the tip of the bow shock is about 
$2.7\times 10^7\,$K ($k_\text{B}\,T\simeq 2.33\,$keV). In contrast, at the bow shock wings, where the shock inclination angle reaches 
approximately $30^\circ$, the downstream temperature has dropped to about $6.3\times 10^6$\,K ($k_\text{B}\,T\simeq 0.54\,$keV). Despite the 
simple assumptions, this value is in remarkably good agreement with the observational data since \cite{Tri:03} measured 
in SQ a post-shock gas temperature of 0.5\,keV.

\subsection{Comparison of the analytical with the numerical model}
\label{sec:comp}
\begin{figure*}
\resizebox{\hsize}{!}{\includegraphics{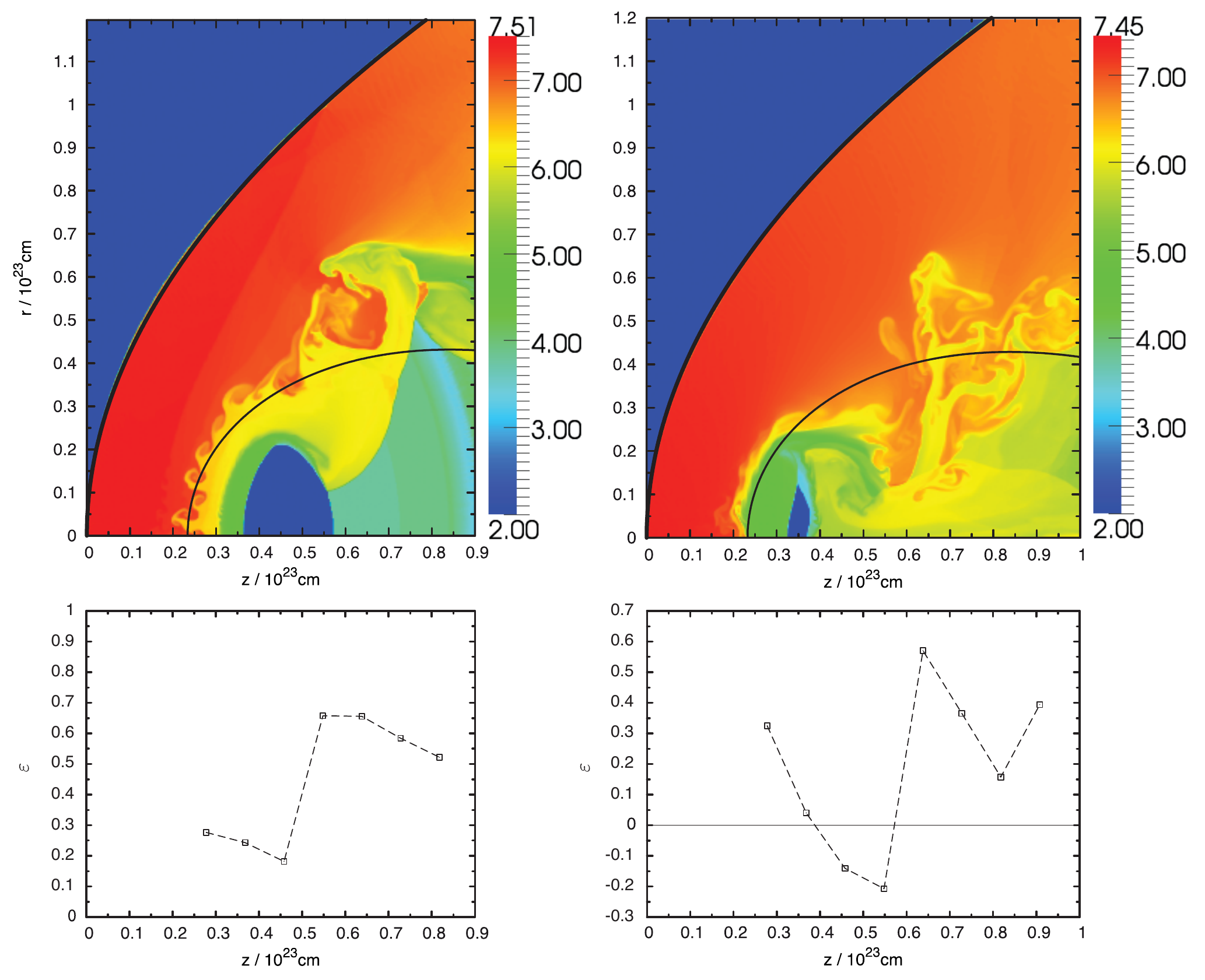}}
\caption{Comparison of numerically (here log temperature in K is colour-coded) and analytically derived results. The thin line represents 
the outer envelope of the galaxy, whereas the thick line represents the bow shock. \textit{Upper left.} Model of the interaction between 
NGC 7318b and the IGM in SQ at the time $t=60\,$Myr ($j=0$, $\hat{r}^2=1.82\,\hat{z}$). \textit{Upper right.} Same as upper left, but at 
at $t=130\,$Myr ($j=0$, $\hat{r}^2=1.81\,\hat{z}$). The \textit{lower panels} give the corresponding percentage errors.}
\label{im:compa1}
\end{figure*}

\begin{figure*}
\resizebox{\hsize}{!}{\includegraphics{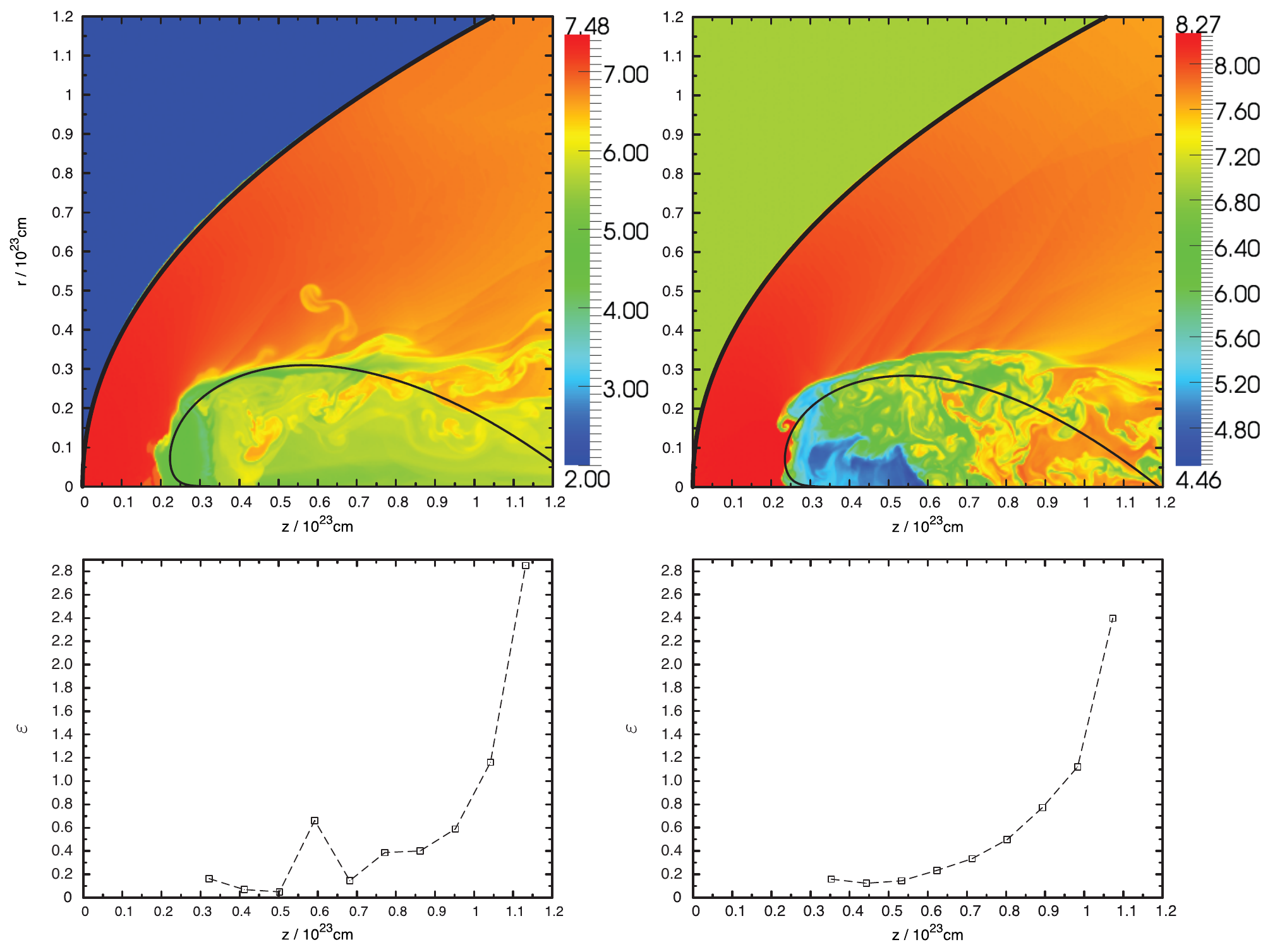}}
\caption{Like Fig.~\ref{im:compa1}. \textit{Upper left.} Model of the interaction between NGC 7318b and the IGM in SQ at the time 
$t=200$\,Myr ($j=0$, $\hat{r}^{2.1}=1.40\,\hat{z}$). \textit{Upper right.} Model of the interaction between an IC 1263-like galaxy and the 
IGM in the galaxy group around IC 1262 at the time $t=130$\,Myr ($j=0$, $\hat{r}^{2.1}=1.39\,\hat{z}$). The upstream Mach number 
here is only $M_{\infty}=5$. The \textit{lower panels} give the corresponding percentage errors.}
\label{im:compa2}
\end{figure*}

In Figs.~\ref{im:compa1} and \ref{im:compa2} our analytical results are superimposed on the numerically derived maps. 
In the lower panels the percentage errors are given. 
We used a shock tracer that scans the computational domain for entropy jumps in order to parametrize the numerically calculated bow 
shocks. In addition to numerical models of the galaxy-IGM interaction in SQ, we also studied the galaxy group around IC 1262. 
\emph{Chandra} and \emph{XMM-Newton} observations have uncovered an extended cool X-ray ridge at the centre of this group \citep{Tri:07}. 
The authors speculate that this peculiar structure may have been created by efficient ram pressure stripping of the bright spiral member 
IC 1263 as it had passed through the group's dense core ($\rho_{\text{IGM}}=8.0\times 10^{-27}\,\text{g\,cm}^{-3}$, 
$T_{\text{IGM}}=2.0\times 10^7\,$K) at a relative velocity of $\sim 1000\,$km\,s$^{-1}$ (translating to a Mach number of about 5). Then, 
gas supersonically streaming behind the bow shock wave may have heated up the stripped ISM to the observed X-ray 
temperatures. A result of our numerical and analytical computation of this interaction is shown in the lower right panel of 
Fig.~\ref{im:compa1}.

It is no surprise that the errors grow with increasing distance from the stagnation point and time, since instabilities, which 
cannot be predicted analytically, begin to grow owing to the increasing influence of shear flows. Moreover, it must be noted that mainly the 
dense and cold ISM phases of the galaxies (blue to light green regions in Figs.~\ref{im:compa1} and \ref{im:compa2}) drive the bow shock 
waves. Therefore these regions are the ones enclosed by the analytical solution. Bearing these aspects in mind, the agreement is 
quite good for all considered scenarios (even for the lower Mach number case). Furthermore, the analytically derived fluid variable maps 
(Fig.~\ref{im:flowfield}) are in good agreement with the numerical calculations (compare the shock layer temperature gradient shown in the 
lower left panel of Fig.~\ref{im:flowfield} with the upper right panel of Fig.~\ref{im:compa1}). The modulus of the percentage error of 
the log temperature immediately behind and in the vicinity of the bow shock wave lies below $1\,\%$. Also the detachment distances are 
determined quite satisfactorily and the modulus of the percentage error ranges from $\sim 5.9\,\%$ to $\sim 38.5\,\%$. A more detailed 
discussion of the flow and comparison to new detailed X-ray data of SQ and the group around IC 1262 is the subject of a forthcoming paper.

Finally, it should be noted that the astrophysical possibility of this analytical method is hardly restricted to 
galaxy-IGM interaction alone; on the contrary, any scenario, such as bow shocks around subclusters, gas bullets, jets (Herbig-Haro 
objects) or individual stars can be covered with it as well.

\section{Conclusion}
\label{conc}
This paper presented an analytical solution technique for the inviscid hypersonic blunt body problem and applied it for the first 
time to the interaction between galaxies and the IGM. The method follows an inverse approach, which means that the shape of the bow shock 
has to be parametrized to be consistent with the galaxy shape. Such an inverse approach has no substantial disadvantage in the 
view of the fact that at high impact velocities the shape of the bow shock becomes similar to the shape of the body. For lower Mach 
numbers we find that the problem can be solved iteratively. First, we assumed the geometric form of the bow shock, then analytically derived 
the shape of the galaxy, and compared it to the relevant parameters (e.g. axes ratio, dimensions). In the next step, the bow shock curve was 
changed, and the body shape recalculated until convergence was obtained. In all cases we tested, this was achieved fairly rapidly. The 
solution, which is valid in the whole flow field, and in particular takes velocity gradients into account along streamlines, is based on two 
main assumptions; first, the density ratio across the shock has to be large, and, second, the pressure at a point $Q$ 
(Fig.~\ref{im:coordsys}) of the disturbed flow field must not to be very small in comparison to the pressure immediately behind the shock in 
the intersection point of the shock surface with its normal through $Q$. In our derivation, heat conduction, viscosity, as well as terms of the order of 
$\chi$ (density ratio) are neglected. It should be stressed that it is not required that the shock layer, which is the area between 
the bow shock wave and the projectile, has to be thin. The presented treatment thus surpasses in sophistication methods like the analytic 
solution for the thin-shell problem of a hypersonic wind interacting with a rigid sphere by \cite{Can:98} and allows
astrophysical problems that involve bow shocks generated by a hypersonic flow to be tackled more realistically.

For a symmetrically assumed galaxy at a zero angle of attack, the stagnation streamline and the stagnation point are along the centreline. 
The stagnation streamline crosses the bow shock at precisely the normal shock point ($\beta=\pi/2$) and hence the entropy of the 
stagnation streamline in the shock layer is maximal \citep{Hay:66}. However, we found that the galaxy need not necessarily be 
axially symmetric oriented with respect to the flow direction. Then the stagnation streamline does not pass through the normal 
portion of the shock wave so it does not coincide with the maximum entropy streamline that always intersects the shock at right 
angles. For all possible (two-dimensional) galaxy inclination angles, however, these two streamlines lie quite close to each other.
Consequently, the presented method is applicable if the $z$-axis of the coordinate system is parallel to the velocity vector of the galaxy 
and if it goes through the stagnation streamline. The galaxy's bow shock halves above and below the $z$-axis then have to be evaluated 
separately.

Our analytical investigation of galaxy-IGM interactions are in good agreement with results of the two-dimensional numerical simulations 
carried out with the (pure) hydrodynamics code {\small VH-1}, which is PPMLR-based and therefore ideally suited to analysing shocks. As 
long as the initially made assumptions are not violated by Mach numbers that are too low, accurate predictions of the fluid quantities in the post-shock flow field, as well as the shock's stand-off distance, are possible. The high potential of the present analytical 
approximation suggests an application to many other astrophysical problems that involve bow shocks.

\begin{acknowledgements}
It is a pleasure to thank Wilhelm Schneider for helpful discussions and Ian R. Stevens for providing essential parts of the used radiative 
cooling algorithm. Use of {\small VH-1}, developed by the numerical astrophysics group at the University of Virginia 
(\texttt{http://wonka.physics.ncsu.edu/pub/VH-1}), is hereby acknowledged. We also thank the anonymous referee for his/her constructive 
suggestions which helped to improve the paper. The calculations were performed on the computers of the Institute for Astronomy, 
University of Vienna (Austria).
\end{acknowledgements}
\bibliographystyle{aa}
\bibliography{12436shock}
\appendix
\section{Derivation of the pressure in the shock layer}
The derivation of equation (\ref{eq:newbuseneu}) in the main text is presented here in detail. 
The momentum equation (\ref{eq:schnmom1}) can be rewritten as
\begin{equation}
\frac{1}{2}\frac{\partial (u^2+v^2)}{\partial\bar{x}}+\frac{P}{\rho}\frac{\partial \ln P}{\partial \bar{x}}=0\,.
\label{eq:momsimple}
\end{equation}
From equations (\ref{eq:uhat}) and (\ref{eq:vhat}) we can estimate the ratio of the velocity components
\begin{equation}
\frac{v^2}{u^2}=\mathcal{O}(\hat{\chi}^2\tan^2\hat{\beta})+\mathcal{O}\left(\frac{v_\text{b}^2}{u_\text{b}^2}\right)\,,
\label{eq:v/u}
\end{equation}
where $u_\text{b}$ and $v_\text{b}$ are the velocity components at the surface of the body. Because the entropy increases with increasing 
shock inclination angle $\beta$, it is valid to write $\hat{S}=\mathcal{O}(S_*)$. Applying approximation (\ref{eq:ass2}), we have 
$\hat{h}=\mathcal{O}(h)$ for the enthalpy. Moreover, it turns out that
\begin{equation}
u=\mathcal{O}(\hat{u})\,,
\end{equation}
where the energy equation (\ref{eq:schnenergy}) has been used. In contrast, since the velocity increases along a streamline with 
decreasing pressure, it holds that $1/u=\mathcal{O}(1/u_*)$. On the other hand, we can infer from the Rankine-Hugoniot jump conditions 
(\ref{eq:uhat})--(\ref{eq:enthbound}) that $1/u=\mathcal{O}(1/\hat{u})$. This is true everywhere in the shock layer except the region, 
where $\cos\beta_*$ is very small compared to $\cos\hat{\beta}$, i.e. where $0\le \psi \le \hat{\psi}$. To include this region 
we have to write
\begin{equation}
\frac{1}{u}=\mathcal{O}\left(\frac{1}{u_\text{b}+[\hat{u}-u_\text{b}]\psi/\hat{\psi}}\right)\,.
\label{eq:1overu}
\end{equation}
An analogous relation can be found for the density, if $h$ and $P$ in
\begin{equation}
\frac{\partial \rho}{\partial \psi}=\left( \frac{\partial \rho}{\partial P}\right)_h \frac{\partial P}{\partial \psi}+\left(\frac{\partial 
\rho}{\partial h}\right)_P \frac{\partial h}{\partial \psi}
\label{eq:thermo}
\end{equation}
are replaced by means of equations (\ref{eq:schnenergy}) and (\ref{eq:schnmom2}). As long as the thermodynamic functions $h$ and $P$ 
retain their order of magnitude, the thermodynamic functions $(\partial \rho/\partial P)_h$ and $(\partial\rho/\partial h)_P$ for a 
gas also do not change their orders of magnitude. The formal integration of the momentum equation (\ref{eq:schnmom2}) with the 
boundary condition $P=\hat{P}$ at $\psi=\hat{\psi}$ yields
\begin{equation}
\begin{split}
\hat{r}^j(\hat{P}-P)&\left[   1 +\mathcal{O}\left(   \left[ \hat{\kappa}+j\frac{\cos\hat{\beta}}{\hat{r}}\right]\frac{1}{\hat{\psi}-\psi}
\int_\psi^{\hat{\psi}}y\, \text{d}\psi\right)\right]\\
&=\hat{\kappa}\int_\psi^{\hat{\psi}}u\,\text{d}\psi+\int_\psi^{\hat{\psi}}\frac{\partial v}{\partial \bar{x}}\,\text{d}\psi\,.
\label{eq:momsol}
\end{split}
\end{equation} 
Furthermore keeping in mind that $v$ is of the order of $u$ everywhere apart from the stagnation region, we see from equation 
(\ref{eq:thermo}), together with (\ref{eq:momsol}), that the order of magnitude of $\partial \rho/\partial \psi$ does not depend on the 
stream function $\psi$, except maybe in the region near the projectile surface, where $u$ can be much 
smaller than $\hat{u}$. Consequently, we obtain
\begin{equation}
\frac{1}{\rho}=\mathcal{O}\left(\frac{1}{\rho_\text{b}+[\hat{\rho}-\rho_\text{b}]\psi/\hat{\psi}}\right)\,.
\label{eq:1overrho}
\end{equation}
With the aid of our results (\ref{eq:1overu}), (\ref{eq:1overrho}), and (\ref{eq:psihat}), equation (\ref{eq:diffy}) yields
\begin{equation}
y=\mathcal{O}\left(\frac{\hat{\chi}\hat{r}}{\cos \hat{\beta}}\frac{1}{u_\text{b}/\hat{u}-\rho_\text{b}/\hat{\rho}} \ln\frac{u_\text{b}/
\hat{u}+[1-u_\text{b}/\hat{u}]\psi/\hat{\psi}}{\rho_\text{b}/\hat
{\rho}+[1-\rho_\text{b}/\hat{\rho}]\psi/\hat{\psi}}\right)\,.
\label{eq:appsol}
\end{equation}
Finally, combining equations (\ref{eq:diffy1}) and (\ref{eq:appsol}) gives 
\begin{equation}
\frac{v_\text{b}}{u_\text{b}}=\mathcal{O}\left(\hat{\chi}\sin\hat{\beta}\ln\frac{\rho_\text{b}}{\hat{\rho}}\right)\,,
\label{eq:appeq}
\end{equation}
which can be combined with equations (\ref{eq:p_hat}) and (\ref{eq:ass2}) to become
\begin{equation}
\frac{v_\text{b}}{u_\text{b}}=\mathcal{O}\left(\hat{\chi}\left[\frac{P_\text{b}}{\rho_\infty U_\infty^2}\right]^{1/2}\ln\frac{\rho_\text
{b}}{\hat{\rho}}\right)\,.
\label{eq:vb/ub}
\end{equation}
The entropy remains constant on streamlines in the post-shock region. It is thus useful to introduce the effective isentropic exponent
\begin{equation}
\gamma_e=\left(\frac{\partial \ln P}{\partial \ln \rho}\right)_S=\frac{\rho}{P}\left(\frac{\partial P}{\partial \rho}\right)_S\,,
\end{equation}
which is not smaller than 1 for any gas. Consequently, the change in relative pressure along a streamline is not smaller than the change 
of relative density, i.e.,
\begin{equation}
\text{d}\rho \le 0:\,\,\,\frac{P}{\rho}=\mathcal{O}\left(\frac{P_*}{\rho_*}\right)=\mathcal{O}(\chi_* U_\infty^2\sin^2\beta_*)\,.
\end{equation} 
On the other hand, the (dimensionless) quantity $P/\rho_\infty U_\infty^2$ cannot be very large for $\text{d}\rho>0$, so for both 
$\text{d}\rho\le 0$ and $\text{d}\rho>0$, we have
\begin{equation}
\frac{P}{\rho}=U_\infty^2\mathcal{O}(\chi_*)\,,
\label{eq:druck/dichte}
\end{equation}
which also holds on the body streamline. Thus equation (\ref{eq:vb/ub}) becomes
\begin{equation}
\frac{v_\text{b}}{u_\text{b}}=\mathcal{O}\left(\left[\hat{\chi}\chi_* \frac{\rho_\text{b}}{\hat{\rho}}\right]^{1/2}\ln\frac{\rho_\text{b}}
{\hat{\rho}}\right)\,,
\end{equation}
which can be simplified by using the assumption (\ref{eq:ass1}) and that $\rho_\text{b}/\hat{\rho}=\mathcal{O}(1)$
\begin{equation}
\frac{v_\text{b}}{u_\text{b}}=\mathcal{O}(\hat{\chi})\,.
\label{eq:vb/ubsimple}
\end{equation}
If we now temporarily exclude the stagnation region (defined by $\tan^2\hat{\chi}\gg1$) from our consideration, equations (\ref{eq:v/u}) 
and (\ref{eq:vb/ubsimple}) suggest neglecting $v^2$ in comparison with $u^2$ in the momentum equation (\ref{eq:momsimple}). Taking the 
isentropy of the flow on streamlines into account, integration of equation (\ref{eq:momsimple}) yields
\begin{equation}
u^2-u_*^2+2\int_{\ln P_*}^{\ln P}\left(\frac{P}{\rho}\right)_{S=S_*} \text{d}(\ln P)=0\,.
\label{eq:newbuse1}
\end{equation}
If we once again exclude the stagnation region, as well as the region characterized by $u\ll \hat{u}$, the first term of equation 
(\ref{eq:newbuse1}) can be rewritten by using
\begin{equation}
U_\infty^2=\mathcal{O}(u^2)\,.
\end{equation}
With the aid of equation (\ref{eq:druck/dichte}) the integral in equation (\ref{eq:newbuse1}) can be evaluated:
\begin{equation}
\int_{\ln P_*}^{\ln P}\left(\frac{P}{\rho}\right)_{S=S_*} \text{d}(\ln P)=U_\infty^2\mathcal{O}\left(\chi_*\ln\frac{P_*}{P}\right)\,.
\end{equation}
Since the gas may expand substantially on the stagnation streamlines, the term $\ln(P_*/P)$ can become very large. For further analysis 
it is advantageous to split the integral in equation (\ref{eq:newbuse1}) into two parts:  
\begin{equation}
\begin{split}
u^2-u_*^2 &+ 2\int_{\ln P_*}^{\ln \hat{P}}\left(\frac{P}{\rho}\right)_{S=S_*} \text{d}(\ln P)\\
&+ 2\int_{\ln \hat{P}}^{\ln P}\left(\frac{P}{\rho}\right)_{S=S_*} \text{d}(\ln P)=0\,,
\end{split}
\label{eq:newbuse2}
\end{equation}
where the second integral can be evaluated by using the assumption (\ref{eq:ass2}) together with equation (\ref{eq:druck/dichte}):
\begin{equation}
\begin{split}
\int_{\ln \hat{P}}^{\ln P}\left(\frac{P}{\rho}\right)_{S=S_*} \text{d}(\ln P)&= U_\infty^2\mathcal{O}\left(\chi_*\ln\frac{\hat{P}}{P}
\right)\\
&= U_\infty^2\mathcal{O}(\chi_*)\,,
\end{split}
\end{equation}
and thus turns out to be negligible. Equation (\ref{eq:newbuse2}) then simplifies to
\begin{equation}
u^2_{(P)}=u_*^2-2\int_{P_*}^{\hat{P}}\left(\frac{1}{\rho}\right)_{S=S_*}\text{d}P+\dots\,,
\end{equation}
or, equivalently,
\begin{equation}
u^2_{(P)}=u_*^2+2[h_*-h(\hat{P},S_*)]+\dots\,,
\label{eq:unewbuse}
\end{equation}
where the well-known relation for the enthalpy $\text{d}h=\text{d}P/\rho$ (for $\text{d}S=0$) has been used so that $h=h(P,S)$ is an 
equation of state of the gas. The subscript $(P)$ implies that this approximation is used only to calculate the pressure $P$. 
Finally, by inserting equation (\ref{eq:unewbuse}), the Newton-Busemann pressure law can be rewritten to yield the important result
\begin{equation}
P=\hat{P}-\frac{\hat{\kappa}}{\hat{r}^j}\int_\psi^{\hat{\psi}}\{u_*^2+2[h_*-h(\hat{P},S_*)]\}^{1/2}\text{d}\psi'\tag{26}\,,
\end{equation}
used in Section \ref{method}.
\end{document}